\documentclass[12pt]{iopart}

\usepackage{iopams}
\expandafter\let\csname equation*\endcsname\relax
\expandafter\let\csname endequation*\endcsname\relax

\usepackage{graphicx}
\usepackage{amssymb,amsfonts,amsmath}
\usepackage{dcolumn}
\usepackage[numbers,square,sort&compress]{natbib}
\usepackage{booktabs}
\usepackage{multirow}
\usepackage{color}
\usepackage{rotating}
\usepackage{pdflscape}

\begin{document}

\title[Joint multifractal analysis based on the partition function approach]{Joint multifractal analysis based on the partition function approach: Analytical analysis, numerical simulation and empirical application}

\author{Wen-Jie Xie$^{1,2,3}$, Zhi-Qiang Jiang$^{1,2}$, Gao-Feng Gu$^{1,2}$, Xiong Xiong$^{4,5}$, Wei-Xing Zhou$^{1,2,6}$}

\address{$^1$ School of Business, East China University of Science and Technology, Shanghai 200237, China} %
\address{$^2$ Research Center for Econophysics, East China University of Science and Technology, Shanghai 200237, China} %
\address{$^3$ Postdoctoral Research Station, East China University of Science and Technology, Shanghai 200237, China}
\address{$^4$ College of Management and Economics, Tianjin University, Tianjin 300072, China}
\address{$^5$ China Center for Social Computing and Analytics, Tianjin University, Tianjin 300072, China}
\address{$^6$ Department of Mathematics, East China University of Science and Technology, Shanghai 200237, China} %
\ead{xxpeter@tju.edu.cn and wxzhou@ecust.edu.cn}

\begin{abstract}
  Many complex systems generate multifractal time series which are long-range cross-correlated. Numerous methods have been proposed to characterize the multifractal nature of these long-range cross correlations. However, several important issues about these methods are not well understood and most methods consider only one moment order. We study the joint multifractal analysis based on partition function with two moment orders, which was initially invented to investigate fluid fields, and derive analytically several important properties. We apply the method numerically to binomial measures with multifractal cross correlations and bivariate fractional Brownian motions without multifractal cross correlations. For binomial multifractal measures, the explicit expressions of mass function, singularity strength and multifractal spectrum of the cross correlations are derived, which agree excellently with the numerical results. We also apply the method to stock market indexes and unveil intriguing multifractality in the cross correlations of index volatilities.
\end{abstract}

\vspace{2pc}
\submitto{\NJP}
\maketitle
% \tableofcontents

\newpage

\section{Introduction}

Measurements of a complex evolving system from different view angles provide us many time series that are usually long-range cross-correlated and exhibit multifractal nature. In turbulent flows, there are velocity field, temperature field and concentration field embedded in the same spatial domain. One can measure these quantities at fixed locations to obtain time series, which are mutually correlated \cite{Antonia-VanAtta-1975-JFM,Meneveau-Sreenivasan-Kailasnath-Fan-1990-PRA}. In financial markets, there are also many pairs that are cross-correlated, such as market index volatilities, price returns of different markets, price returns of different equities, different quantities of a same equity \cite{Lin-2008-PA,Podobnik-Horvatic-Petersen-Stanley-2009-PNAS,SiqueirJr-Stosic-Bejan-Stosic-2010-PA,Wang-Wei-Wu-2010-PA,He-Chen-2011-CSF,Zhou-2012-QF,Zhou-2012-NJP,Zhuang-Wei-Zhang-2014-PA}. Moreover, examples come from very diverse fields, including agronomy \cite{Kravchenko-Bullock-Boast-2000-AJ,Zeleke-Si-2004-AJ}, seismic data \cite{Shadkhoo-Jafari-2009-EPJB}, meteorology \cite{JimenezHornero-JimenezHornero-deRave-PavonDominguez-2010-EMA,JimenezHornero-PavonDominguez-deRave-ArizaVillaverde-2010-AR,Shen-Li-Si-2015-PA}, medical science \cite{Lin-Sharif-2010-Chaos,Ghosh-Dutta-Chakraborty-2014-CSF}, geophysics \cite{Hajian-Movahed-2010-PA}, transportation \cite{Xu-Shang-Kamae-2010-ND,Zhao-Shang-Lin-Chen-2011-PA,Zebende-daSilva-Filho-2011-PA}, to list a few.

To extract the joint multifractality between a pair of multifractal time series, a variety of methods have been developed, such as the MF-X-PF method that performs joint multifractal analysis \cite{Antonia-VanAtta-1975-JFM,Meneveau-Sreenivasan-Kailasnath-Fan-1990-PRA,Schmitt-Schertzer-Lovejoy-Brunet-1996-EPL,Xu-Antonia-Rajagopalan-2000-EPL,Xu-Antonia-Rajagopalan-2007-EPL,Wang-Shang-Ge-2012-Fractals} based on the partition function approach \cite{Halsey-Jensen-Kadanoff-Procaccia-Shraiman-1986-PRA}, the MF-X-DFA method that conducts multifractal detrended cross-correlation analysis \cite{Zhou-2008-PRE} based on the detrended fluctuation analysis \cite{Peng-Buldyrev-Havlin-Simons-Stanley-Goldberger-1994-PRE,Kantelhardt-KoscielnyBunde-Rego-Havlin-Bunde-2001-PA}, multifractal detrended fluctuation analysis
\cite{CastroESilva-Moreira-1997-PA,Weber-Talkner-2001-JGR,Kantelhardt-Zschiegner-KoscielnyBunde-Havlin-Bunde-Stanley-2002-PA}, and the detrended cross-correlation analysis \cite{Jun-Oh-Kim-2006-PRE,Podobnik-Stanley-2008-PRL,Podobnik-Horvatic-Petersen-Stanley-2009-PNAS,Horvatic-Stanley-Podobnik-2011-EPL,Kristoufek-2013-EPJB,Kristoufek-2014a-PA,Ying-Shang-2014-Fractals,Kristoufek-2015-PRE}, the MF-X-DMA method \cite{Jiang-Zhou-2011-PRE} that carries out the multifractal detrended cross-correlation analysis based on the detrending moving-average analysis \cite{Vandewalle-Ausloos-1998-PRE,Alessio-Carbone-Castelli-Frappietro-2002-EPJB,Carbone-Castelli-2003-SPIE,Carbone-Castelli-Stanley-2004-PA,Carbone-Castelli-Stanley-2004-PRE,Varotsos-Sarlis-Tanaka-Skordas-2005-PRE,Xu-Ivanov-Hu-Chen-Carbone-Stanley-2005-PRE,Arianos-Carbone-2007-PA,Bashan-Bartsch-Kantelhardt-Havlin-2008-PA,Arianos-Carbone-2009-JSM,Carbone-2009-IEEE} and multifractal detrending moving-average analysis \cite{Gu-Zhou-2010-PRE,He-Chen-2011b-PA}, the multifractal height cross-correlation analysis (MF-HXA) method \cite{Kristoufek-2011-EPL}, the multiscale multifractal detrended cross-correlation analysis (MM-DCCA) \cite{Shi-Shang-Wang-Lin-2014-PA}, and the MF-DPXA method  \cite{Qian-Liu-Jiang-Podobnik-Zhou-Stanley-2015-PRE} that generalizes the detrended partial cross-correlation analysis \cite{Liu-2014,Yuan-Fu-Zhang-Piao-Xoplaki-Luterbacher-2015-SR} in which the partial correlation is considered. Properly designed statistical tests can be used to quantify these cross correlations \cite{Podobnik-Grosse-Horvatic-Ilic-Ivanov-Stanley-2009-EPJB,Zebende-2011-PA,Podobnik-Jiang-Zhou-Stanley-2011-PRE}.

The joint multifractal analysis is a classic method and has been applied to study the joint multifractal nature between different pairs of time series recorded in natural and social sciences \cite{Kravchenko-Bullock-Boast-2000-AJ,Zeleke-Si-2004-AJ,Lin-2008-PA,JimenezHornero-JimenezHornero-deRave-PavonDominguez-2010-EMA,Lin-Sharif-2010-Chaos,JimenezHornero-PavonDominguez-deRave-ArizaVillaverde-2010-AR,Wang-Shang-Ge-2012-Fractals}. Due to its elegant geometric nature, many important properties can be derived, which is however very difficult in the frameworks of other methods mentioned above. For instance, although there is numerical evidence and analytical results for the relationship between the cross-multifractal spectrum $f_{xy}(\alpha_x,\alpha_y)$ and the multifractal spectra $f_x(\alpha_x)$ and $f_y(\alpha_y)$ of individual time series \cite{Podobnik-Stanley-2008-PRL,Zhou-2008-PRE,Jiang-Zhou-2011-PRE,Wang-Shang-Ge-2012-Fractals,Kristoufek-2015-PA}, the problem is not solved. Moreover, the original MF-X-PF method is important because it handles moments with two different orders, while recent methods for multifractal cross-correlation analysis focus only on one order.

In this work, we recover the uni-order MF-X-PF method \cite{Wang-Shang-Ge-2012-Fractals} and propose a direct determination approach for the multifractal spectrum using the idea from the bi-order MF-X-PF framework \cite{Meneveau-Sreenivasan-Kailasnath-Fan-1990-PRA}. Based on this framework, we are able to derive important geometric properties of the uni-order MF-X-PF method. We perform numerical simulations using different mathematical models and explain the results of multifractal binomial measures analytically. Finally, we apply the bi-order MF-X-PF method to stock market indices.

\section{Joint multifractal analysis based on partition function approach}
\label{S2:MF-X-PF}

In this section, we first present the joint multifractal analysis based on partition function approach with two moment orders \cite{Meneveau-Sreenivasan-Kailasnath-Fan-1990-PRA}, abbreviated MF-X-PF$(p,q)$, and then derive the uni-order method MF-X-PF$(q)$ that was independently proposed recently \cite{Wang-Shang-Ge-2012-Fractals}. Although the joint partition function $\chi_{xy}(q,s)$ of the uni-order method can be directly recovered from the joint partition function $\chi_{xy}(p,q,s)$ of the bi-order method by posing $p=q$, we will show that the nexus between the multifractal properties of the two methods is not obvious, which is caused by the application of the steepest descent approach.

\subsection{MF-X-PF$(p,q)$}

Based on the box-counting idea, the geometric support is partitioned into boxes of size $s$. We consider two integrated measures $m_x(s,t)$ and $m_y(s,t)$ in the $t$-th box. The local singularity strengths $\alpha_x$ and $\alpha_y$ are defined according to the following relationships:
\begin{subequations}
\begin{equation}
  m_x(s,t) \sim s^{\alpha_x},
  \label{Eq:MF-X-PF:mx:s:alphax}
\end{equation}
and
\begin{equation}
  m_y(s,t) \sim s^{\alpha_y}.
  \label{Eq:MF-X-PF:my:s:alphay}
\end{equation}
\label{Eq:MF-X-PF:mxy:s:alpha}
\end{subequations}
Let $N_s(\alpha_x,\alpha_y)$ denote the number of boxes of size $s$ needed to cover the set of points in which the singularity strengths are around $\alpha_x$ and $\alpha_y$ with bands $d\alpha_x$ and $d\alpha_y$. Hence, the fractal dimension of the set is determined according to \cite{Mandelbrot-1983}
\begin{equation}
  N_s(\alpha_x,\alpha_y) \sim s^{-f_{xy}(\alpha_x,\alpha_y)},
  \label{Eq:MF-X-PF:Ns:s:f:alpha:xy}
\end{equation}
in which $f_{xy}(\alpha_x,\alpha_y)$ is the joint distribution of the two singularity strengths \cite{Meneveau-Sreenivasan-Kailasnath-Fan-1990-PRA} or the joint multifractal spectrum.

We consider the joint partition function
\begin{equation}
  \chi_{xy}(p,q,s)= \sum_t\left[m_x(s,t)\right]^{p/2}[m_y(s,t)]^{q/2}.
  \label{Eq:MF-X-PF:pq:chi:s}
\end{equation}
This definition is slightly different from that in Ref.~\cite{Meneveau-Sreenivasan-Kailasnath-Fan-1990-PRA}, in which the orders are $p$ and $q$ rather than $p/2$ and $q/2$. In this setting, we recover the traditional partition function when $m_x=m_y$ and $p=q$ \cite{Halsey-Jensen-Kadanoff-Procaccia-Shraiman-1986-PRA}. The joint mass exponent function $\tau_{xy}(p,q)$ can be obtained from the following relation
\begin{equation}
  \chi_{xy}(p,q,s) \sim s^{\tau_{xy}(p,q)}.
  \label{Eq:MF-X-PF:pq:tauxy}
\end{equation}
In practice, for a given pair $(p,q)$, we compute $\chi_{xy}(p,q,s)$ for a various of box sizes $s$ and perform linear regression of $\ln\chi_{xy}(p,q,s)$ against $\ln{s}$ in a proper scaling range to obtain $\tau_{xy}(p,q)$.
%
%If $D_x(p)$ and $D_y(q)$ are the mass exponents for individual measures $x$ and $y$, we have
%\begin{subequations}
%\begin{equation}
%  \chi_{xy}(p,0,s) =\sum_t|m_x(s,t)|^{p/2}|m_y(s,t)|^{0/2}\sim s^{(p/2-1)D_x(p/2)},
%  \label{Eq:MF-X-PF:Dx:p}
%\end{equation}
%and
%\begin{equation}
%  \chi_{xy}(0,q,s) = \sum_t|m_x(s,t)|^{0/2}|m_y(s,t)|^{q/2} \sim s^{(q/2-1)D_y(q/2)}.
%  \label{Eq:MF-X-PF:Dy:q}
%\end{equation}
%\end{subequations}
%Together with Eq.~(\ref{Eq:MF-X-PF:pq:tauxy}), we have
%\begin{subequations}
%\begin{equation}
%  \tau_{xy}(p,0) = (p/2-1)D_x(p/2),
%  \label{Eq:MF-X-PF:tau:p:0:Dx}
%\end{equation}
%and
%\begin{equation}
%  \tau_{xy}(0,q) = (q/2-1)D_y(q/2).
%  \label{Eq:MF-X-PF:tau:0:1:Dy}
%\end{equation}
%\end{subequations}

We insert the two relations in Eq.~(\ref{Eq:MF-X-PF:mxy:s:alpha}) into the joint partition function, rewrite the sum into a double integral over $\alpha_x$ and $\alpha_y$, and then apply the steepest descent approach to estimate the integral at small $s$ values, which leads to
\begin{equation}
  \tau_{xy}(p,q) = p\alpha_x/2+q\alpha_y/2 - f_{xy}(\alpha_x,\alpha_y),
  \label{Eq:MF-X-PF:tau:alpha:f}
\end{equation}
where
\begin{subequations}
\begin{equation}
  {\partial f_{xy}(\alpha_x,\alpha_y)}/{\partial \alpha_x} =  {p}/{2},
  \label{Eq:MF-X-PF:df:alphax:p}
\end{equation}
and
\begin{equation}
  {\partial f_{xy}(\alpha_x,\alpha_y)}/{\partial \alpha_y} =  {q}/{2}.
  \label{Eq:MF-X-PF:df:alphay:q}
\end{equation}
\end{subequations}

Taking partial derivative of Eq.~(\ref{Eq:MF-X-PF:tau:alpha:f}) over $p$, we have
\begin{equation}
  {\partial\tau_{xy}(p,q)}/{\partial p} = {\alpha_x}/{2}.
%  \frac{\partial\tau_{xy}(p,q)}{\partial p}
%    =  \frac{\alpha_x}{2} +\frac{p}{2}\frac{\partial\alpha_{x}}{\partial p} +\frac{q}{2}\frac{\partial\alpha_{y}}{\partial p}
%      -\frac{\partial f}{\partial \alpha_x}\frac{\partial\alpha_{x}}{\partial p}
%      -\frac{\partial f}{\partial \alpha_y}\frac{\partial\alpha_{y}}{\partial p}
%    =  \frac{\alpha_x}{2}.
\end{equation}
Similar derivation can be done over $q$ and one can obtain the double Legendre transforms
\begin{subequations}
\begin{equation}
  \alpha_x = 2{\partial \tau_{xy}(p,q)}/{\partial p},
  \label{Eq:MF-X-PF:alphax:tau}
\end{equation}
\begin{equation}
  \alpha_y = 2{\partial \tau_{xy}(p,q)}/{\partial q},
  \label{Eq:MF-X-PF:alphay:tau}
\end{equation}
\begin{equation}
  f_{xy}(\alpha_x,\alpha_y) =  p\alpha_x(p,q)/2+q\alpha_y(p,q)/2 -\tau_{xy}(p,q).
  \label{Eq:MF-X-PF:f:alpha}
\end{equation}
\label{Eq:MF-X-PF:pq:Legendre}
\end{subequations}
Therefore, after obtaining $\tau_{xy}(p,q)$, we can numerically determine $\alpha_x$ using Eq.~(\ref{Eq:MF-X-PF:alphax:tau}), $\alpha_y$ using Eq.~(\ref{Eq:MF-X-PF:alphay:tau}), and $f_{xy}(\alpha_x,\alpha_y)$ using Eq.~(\ref{Eq:MF-X-PF:f:alpha}).

From the canonical perspective, one can obtain the $f_{xy}(\alpha_x,\alpha_y)$ function directly \cite{Chhabra-Jensen-1989-PRL,Chhabra-Meneveau-Jensen-Sreenivasan-1989-PRA,Meneveau-Sreenivasan-Kailasnath-Fan-1990-PRA}. Defining the canonical measures as follows
\begin{equation}
  \mu_{xy}(p,q,s,t) = \frac{[m_x(s,t)]^{p/2}[m_y(s,t)]^{q/2}}{\sum_t [m_x(s,t)]^{p/2}[m_y(s,t)]^{q/2}}, % \chi_{xy}(p,q,s)
  \label{Eq:MF-X-PF:pq:muxy}
\end{equation}
the two singularity strengths $\alpha_x(p,q)$ and $\alpha_y(p,q)$ and the joint multifractal spectrum $f_{xy}(p,q)$ can be computed by linear regressions in log-log scales using the following equations:
\begin{subequations}
\begin{equation}
  {\alpha_x(p,q)} = \lim_{s\to0} \frac{\sum_t \mu_{xy}(p,q,s,t) \ln{m_x(s,t)}}{\ln{s}},
  \label{Eq:MF-X-PF:pq:alphax:mu:mx}
\end{equation}
\begin{equation}
  {\alpha_y(p,q)} = \lim_{s\to0}\frac{\sum_t \mu_{xy}(p,q,s,t) \ln{m_y(s,t)}}{\ln{s}},
  \label{Eq:MF-X-PF:pq:alphay:mu:my}
\end{equation}
\begin{equation}
  {f_{xy}(p,q)}  = \lim_{s\to0}\frac{\sum_t \mu_{xy}(p,q,s,t) \ln\left[\mu_{xy}(p,q,s,t)\right]}{\ln{s}}.
  \label{Eq:MF-X-PF:pq:fpq:mu}
\end{equation}
\label{Eq:MF-X-PF:pq:Direct}
\end{subequations}
The joint mass exponent function can be obtained by using Eq.~(\ref{Eq:MF-X-PF:tau:alpha:f}).

\subsection{MF-X-PF$(q)$}

The multifractal cross-correlation analysis based on statistical moments (MFSMXA) proposed in Ref.~\cite{Wang-Shang-Ge-2012-Fractals} is actually a special case of MF-X-PF$(p,q)$ when $p=q$. We call it MF-X-PF$(q)$ here for consistency. In this case, we have
\begin{equation}
  \chi_{xy}(q,s)= \sum_t\left[m_x(s,t)m_y(s,t)\right]^{q/2}  \sim s^{\tau_{xy}(q)},
  \label{Eq:MF-X-PF:q:chi:s}
\end{equation}
in which $\tau_{xy}(q,q)\triangleq \tau_{xy}(q)$. Applying the method of steepest descent, Eq.~(\ref{Eq:MF-X-PF:tau:alpha:f}) becomes
\begin{equation}
  \tau_{xy}(q) = q(\alpha_x+\alpha_y)/2 - f_{xy}(\alpha_x,\alpha_y),
  \label{Eq:MF-X-PF:q:tau:ax:ay:f}
\end{equation}
where
\begin{equation}
  {\partial f_{xy}(\alpha_x,\alpha_y)}/{\partial \alpha_x} = {\partial f_{xy}(\alpha_x,\alpha_y)}/{\partial \alpha_y} = q/2.
  \label{Eq:MF-X-PF:q:df:dalphax:q}
\end{equation}
Taking derivative of Eq.~(\ref{Eq:MF-X-PF:q:tau:ax:ay:f}) over $q$ and using Eq.~(\ref{Eq:MF-X-PF:q:df:dalphax:q}), we have
\begin{equation}
  \frac{d\tau_{xy}(q)}{dq} =  \frac{\alpha_x}{2} +\frac{q}{2}\frac{d\alpha_{x}}{dq} +\frac{\alpha_y}{2} + \frac{q}{2}\frac{d\alpha_{y}}{dq}
    -\frac{\partial f_{xy}}{\partial \alpha_x}\frac{d\alpha_{x}}{dq}
    -\frac{\partial f_{xy}}{\partial \alpha_y}\frac{d\alpha_{y}}{dq}
    =  \frac{\alpha_x+\alpha_y}{2}.
  \label{Eq:MF-X-PF:q:dtau:dq:alphax:alphay}
\end{equation}
Defining that
\begin{equation}
  \alpha_{xy} \triangleq [\alpha_{x}(q) + \alpha_{y}(q)]/2,
  \label{Eq:MF-X-PF:q:alpha:alphax:alphay}
\end{equation}
Eq.~(\ref{Eq:MF-X-PF:q:dtau:dq:alphax:alphay}) and Eq.~(\ref{Eq:MF-X-PF:q:tau:ax:ay:f}) can be rewritten as follows
\begin{subequations}
\begin{equation}
  \alpha_{xy} = d\tau_{xy}(q)/dq,
  \label{Eq:MF-X-PF:q:dtau:dq:alpha}
\end{equation}
\begin{equation}
  f_{xy}(\alpha_{xy}(q)) =  q \alpha_{xy}(q) -\tau_{xy}(q),
  \label{Eq:MF-X-PF:q:f:alpha}
\end{equation}
\label{Eq:MF-X-PF:q:Legendre}
\end{subequations}
where $f_{xy}(\alpha_{xy})\triangleq f_{xy}(\alpha_x,\alpha_y)$. We notice that Eq.~(\ref{Eq:MF-X-PF:q:Legendre}) has the same form of the Legendre transform \cite{Halsey-Jensen-Kadanoff-Procaccia-Shraiman-1986-PRA}.

Because $\alpha_x=d\tau_x(q)/dq$ and $\alpha_y=d\tau_y(q)/dq$ \cite{Halsey-Jensen-Kadanoff-Procaccia-Shraiman-1986-PRA}, it is easy to verify that the following relationship
\begin{equation}
  \tau_{xy}(q)= [\tau_x(q) + \tau_y(q)]/2 + C
  \label{Eq:MF-X-PF:q:tauxy:taux:tauy:c}
\end{equation}
satisfies Eq.~(\ref{Eq:MF-X-PF:q:dtau:dq:alpha}), where $C$ is a constant. According to Eq.~(\ref{Eq:MF-X-PF:q:chi:s}), we have $\chi(0,s) \sim s^{\tau_{xy}(0)} \sim s^{-D_0}$, which is used to measure the fractal dimension of the geometric support. It follows that
\begin{equation}
  \tau_{xy}(0)= -D_0 = -1.
  \label{Eq:MF-X-PF:q:tauxy:q=0}
\end{equation}
Combining Eqs.~(\ref{Eq:MF-X-PF:q:tauxy:taux:tauy:c}) and (\ref{Eq:MF-X-PF:q:tauxy:q=0}) and using $\tau_x(0)=\tau_y(0)=-1$, we have $C=0$ and thus
\begin{equation}
  \tau_{xy}(q)= [\tau_x(q) + \tau_y(q)]/2.
  \label{Eq:MF-X-PF:q:tauxy:taux:tauy}
\end{equation}
Inserting Eq.~(\ref{Eq:MF-X-PF:q:alpha:alphax:alphay}) and Eq.~(\ref{Eq:MF-X-PF:q:tauxy:taux:tauy}) into Eq.~(\ref{Eq:MF-X-PF:q:f:alpha}), we obtain that
\begin{equation}
  f_{xy}(q)= [f_x(q) + f_y(q)]/2.
  \label{Eq:MF-X-PF:q:fxy:fx:fy}
\end{equation}
We note that Eq.~(\ref{Eq:MF-X-PF:q:tauxy:taux:tauy}) and Eq.~(\ref{Eq:MF-X-PF:q:fxy:fx:fy}) still hold when $D_0\neq1$. In this case, we use $\tau_{xy}(0)=\tau_x(0)=\tau_y(0)= -D_0$ to conduct the derivation. These relations were observed numerically using the MF-X-DFA method \cite{Zhou-2008-PRE}, the MF-X-DMA method \cite{Jiang-Zhou-2011-PRE} and the MF-X-PF$(q)$ method \cite{Wang-Shang-Ge-2012-Fractals}.

As shown in Eq.~(\ref{Eq:MF-X-PF:q:chi:s}), the problem is to handle a measure $[m_x(s,t)m_y(s,t)]^{1/2}$. From the canonical perspective, we can obtain the $f_{xy}(\alpha_{xy})$ function directly \cite{Chhabra-Jensen-1989-PRL,Chhabra-Meneveau-Jensen-Sreenivasan-1989-PRA,Meneveau-Sreenivasan-Kailasnath-Fan-1990-PRA}. We can define the canonical measures
\begin{equation}
  \mu_{xy}(q,s,t) = \frac{[m_x(s,t)m_y(s,t)]^{q/2}}{\sum_t [m_x(s,t)m_y(s,t)]^{q/2}}. % \chi_{xy}(p,q,s)
  \label{Eq:MF-X-PF:q:mu:p:q}
\end{equation}
The two singularity strengths $\alpha_x(p)$ and $\alpha_x(p)$ and the joint multifractal spectrum $f_{xy}(p,q)$ can be computed by linear regressions in log-log scales using the following equations:
\begin{subequations}
\begin{equation}
  \alpha_{xy}(q) = \lim_{s\to0} \frac{\sum_t \mu_{xy}(q,s,t) \ln{[m_x(s,t)m_y(s,t)]^{1/2}}}{\ln{s}}
      = \frac{\alpha_{x}(q)+\alpha_{y}(q)}{2},
  \label{Eq:MF-X-PF:q:alpha:mu:mx}
\end{equation}
where Eq.~(\ref{Eq:MF-X-PF:pq:alphax:mu:mx}) and Eq.~(\ref{Eq:MF-X-PF:pq:alphay:mu:my}) are used in the second equality, and
\begin{equation}
  f_{xy}(\alpha_{xy}(q))  = \lim_{s\to0}\frac{\sum_t \mu_{xy}(q,s,t) \ln\left[\mu_{xy}(q,s,t)\right]}{\ln{s}}.
  \label{Eq:MF-X-PF:q:fpq:mu}
\end{equation}
\end{subequations}
The joint mass exponent function can be obtained by using Eq.~(\ref{Eq:MF-X-PF:q:f:alpha}).

\section{Joint multifractal analysis of binomial measures}

\subsection{Numerical analysis applying MF-X-PF$(p,q)$}

\begin{figure}[tb]
\centering
\includegraphics[width=0.96\linewidth]{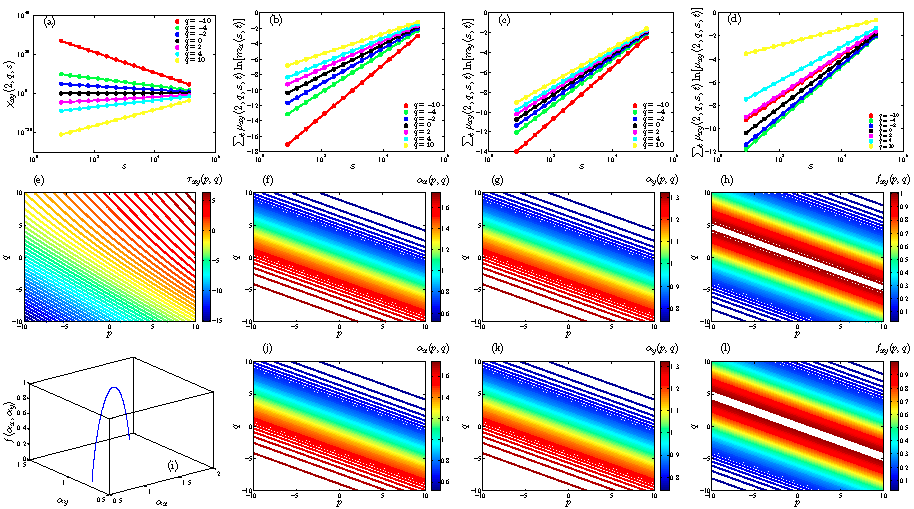}
  \caption{Joint multifractal analysis of two binomial measures with $p_x=0.3$ and $p_y=0.4$ based on the bi-order MF-X-PF$(p,q)$ method. (a) Power-law dependence of $\chi_{xy}(p,q,s)$ on box size $s$ for different $q$ with fixed $p=2$. (b) Linear dependence of $\sum_t\mu_{xy}(2,q,s,t)\ln[m_{x}(s,t)]$ against $\ln{s}$ for different $q$ with fixed $p=2$. (c) Linear dependence of $\sum_t\mu_{xy}(2,q,s,t)\ln[m_{y}(s,t)]$ against $\ln{s}$ for different $q$ with fixed $p=2$. (d) Linear dependence of $\sum_t\mu_{xy}(2,q,s,t)\ln[\mu_{xy}(2,q,s,t)]$ against $\ln{s}$ for different $q$ with fixed $p=2$. (e-i) Mass exponent function $\tau_{xy}(p,q)$, singularity functions $\alpha_x(p,q)$ and $\alpha_y(p,q)$, and multifractal spectra $f_{xy}(p,q)$ and $f_{xy}(\alpha_x, \alpha_y)$ obtained from (a). (j) Singularity function $\alpha_x(p,q)$ obtained from (b). (k) Singularity function $\alpha_y(p,q)$ obtained from (c). (l) Multifractal spectrum $f_{xy}(p,q)$ obtained from (d). }
  \label{Fig:MF-X-PF:pq:pmodel}
\end{figure}

We perform joint multifractal analysis numerically of two binomial measures \cite{Meneveau-Sreenivasan-1987-PRL}. We use $p_x=0.3$ and $p_y=0.4$ and generate two binomial measures of length $2^{20}$. Figure \ref{Fig:MF-X-PF:pq:pmodel}(a) shows on log-log scales the dependence of $\chi_{xy}(p,q,s)$ against box size $s$ for different $q$ with fixed $p=2$. It is obvious that the curves for different $q$ exhibit excellent power law relationships. The power-law exponents obtained by linear regressions of $\ln\chi_{xy}(p,q,s)$ against $\ln s$ are estimates of the mass exponents $\tau_{xy}(p,q)$, whose contour plot is shown in Fig.~\ref{Fig:MF-X-PF:pq:pmodel}(e). We find that $\tau_{xy}(p,q)$ increases with $p$ and $q$. Adopting the double Legendre transform in Eq.~(\ref{Eq:MF-X-PF:pq:Legendre}), we obtain numerically the singularity functions $\alpha_x(p,q)$ and $\alpha_y(p,q)$ and the multifractal spectrum $f_{xy}(p,q)$, whose contour plots are illustrated in Fig.~\ref{Fig:MF-X-PF:pq:pmodel}(f-h) respectively. We find that $\alpha_x(p,q)$ and $\alpha_y(p,q)$ are decreasing functions of $p$ and $q$, while $f_{xy}(p,q)$ has a saddle shape. An intriguing feature is that the contour lines are parallel to each other for $\alpha_x(p,q)$, $\alpha_y(p,q)$ and $f_{xy}(p,q)$. Figure \ref{Fig:MF-X-PF:pq:pmodel}(i) plots the singularity spectrum $f_{xy}(\alpha_x, \alpha_y)$, which is not a surface but a curve.

We also calculate the multifractal functions using the direct determination approach presented in Eq.~(\ref{Eq:MF-X-PF:pq:Direct}) for comparison. In Fig.~\ref{Fig:MF-X-PF:pq:pmodel}(b) to Fig.~\ref{Fig:MF-X-PF:pq:pmodel}(d), we illustrate respectively the linear dependence of $\sum_t\mu_{xy}(2,q,s,t)\ln[m_{x}(s,t)]$, $\sum_t\mu_{xy}(2,q,s,t)\ln[m_{y}(s,t)]$ and $\sum_t\mu_{xy}(2,q,s,t)\ln[\mu_{xy}(2,q,s,t)]$ against $\ln{s}$ for different $q$ with fixed $p=2$. The singularity strength functions $\alpha_x(p,q)$ and $\alpha_y(p,q)$  and the multifractal spectrum $f_{xy}(p,q)$ are computed from the slopes of the lines in these three plots. The corresponding contour plots are presented in Fig.~\ref{Fig:MF-X-PF:pq:pmodel}(j) to Fig.~\ref{Fig:MF-X-PF:pq:pmodel}(l), which are the same as those in Fig.~\ref{Fig:MF-X-PF:pq:pmodel}(f) to Fig.~\ref{Fig:MF-X-PF:pq:pmodel}(h). The numerical results presented in to Fig.~\ref{Fig:MF-X-PF:pq:pmodel} can be derived analytically.

\subsection{Analytical results for MF-X-PF$(p,q)$}

Let us start with two multifractal binomial measures of length $2^L$. Consider two integrated measures $m_x(s,t)$ and $m_y(s,t)$ in boxes of size $s=2^{l}$. There are $n$ types of boxes whose integrated measures are different, in which
\begin{equation}
  n=L-l=L-\frac{\ln s}{\ln 2}.
  \label{Eq:MF-X-PF:pq:n}
\end{equation}
For the $t$-the box, we have
\begin{subequations}
\begin{equation}
  m_x(s,t)  = m_x(2^{n+1},t) =p_x^k(1-p_x)^{n-k},
  \label{Eq:MF-X-PF:analytic:mx}
\end{equation}
\begin{equation}
  m_y(s,t)  = m_y(2^{n+1},t) =p_y^k(1-p_y)^{n-k},
  \label{Eq:MF-X-PF:analytic:my}
\end{equation}
\end{subequations}
where $k\in\{1,...,n\}$.
% \begin{equation}
% \ln m_x(s,t)  =  k[\ln p_x-\ln (1-p_x)]+n\ln (1-p_x)
%  \label{Eq:MF-X-PF:analytic:lnmx}
%\end{equation}
%\begin{equation}
% \ln m_y(s,t)  =  k[\ln p_y-\ln (1-p_y)]+n\ln (1-p_y)
%  \label{Eq:MF-X-PF:analytic:lnmy}
%\end{equation}
It follows that
\begin{equation}
  k = \frac{\ln m_y(s,t)-n\ln(1-p_y)}{\ln p_y-\ln (1-p_y)}.
  \label{Eq:MF-X-PF:analytic:k}
\end{equation}
Inserting Eq.~(\ref{Eq:MF-X-PF:analytic:k}) into Eq.~(\ref{Eq:MF-X-PF:analytic:mx}), we get
\begin{equation}
  m_x(s,t) =C(s) \left[m_y(s,t)\right]^{\beta} = e^{-\gamma L}s^{{\gamma}/{\ln 2}} \left[m_y(s,t)\right]^{\beta},
  \label{Eq:MF-X-PF:analytic:mxmy}
\end{equation}
where
%\begin{equation}
%  C(s)=e^{-n\gamma} = s^{{\gamma}/{\ln 2}}e^{-\gamma L},
%  \label{Eq:MF-X-PF:analytic:Cs}
%\end{equation}
\begin{equation}
  \beta=\frac{\ln p_x-\ln (1-p_x)}{\ln p_y-\ln (1-p_y)}
  \label{Eq:MF-X-PF:analytic:beta}
\end{equation}
and
\begin{equation}
  \gamma=\beta\ln (1-p_y) -\ln (1-p_x).
  \label{Eq:MF-X-PF:analytic:gamma}
\end{equation}
Note that $\beta$ and $\gamma$ depend only on $p_x$ and $p_y$. When $p_x+p_y=1$, we have $\beta=-1$. When $p_x=p_y$, we have $\beta=1$ and $C(s)=1$. When both $p_x$ and $p_y$ are greater than 0.5 or less than 0.5, that is, $(p_x-0.5)(p_y-0.5)>0$, we have $\beta>0$; Otherwise, when $(p_x-0.5)(p_y-0.5)<0$, we have $\beta<0$.

Combining Eq.~(\ref{Eq:MF-X-PF:pq:chi:s}) and Eq.~(\ref{Eq:MF-X-PF:analytic:mxmy}), we obtain
\begin{equation}
  \chi_{xy}(p,q,s)  = C(s)^{p/2}\sum_t\left[m_y(s,t)\right]^{Q}
  \label{Eq:MF-X-PF:analytic:chi:s}
\end{equation}
%\begin{equation}
% \begin{aligned}
%  \chi_{xy}(p,q,s) &= \sum_t\left[m_x(s,t)\right]^{p/2}[m_y(s,t)]^{q/2}\\
%  &= \sum_tC(s)^{p/2}\left[m_y(s,t)\right]^{\beta p/2}[m_y(s,t)]^{q/2}\\
%  &= C(s)^{p/2}\sum_t\left[m_y(s,t)\right]^{\beta p/2+q/2}
%  \label{Eq:MF-X-PF:analytic:chi:p:q:s}
%  \end{aligned}
%\end{equation}
where
\begin{equation}
 Q = \beta p/2+q/2
  \label{Eq:MF-X-PF:analytic:chi:p:q:c}.
\end{equation}
Because $m_y$ is a multifractal measure, we have
\begin{equation}
  \sum_t\left[m_y(s,t)\right]^{Q}\sim s^{\tau_y(Q)}
\end{equation}
where $\tau_y(Q)$ has an analytical expression \cite{Halsey-Jensen-Kadanoff-Procaccia-Shraiman-1986-PRA}:
\begin{equation}
 \tau_y(Q) =- \ln[p_y^{Q}+(1-p_y)^{Q}]/\ln{2}.
 \label{Eq:MF-X-PF:analytic:tauy}
\end{equation}
The joint partition function can be rewritten as follows
\begin{equation}
  \chi_{xy}(p,q,s) \sim s^{\frac{p\gamma}{2\ln 2}}e^{\frac{-p\gamma L}{2}}s^{\tau_y(Q)}.
  \label{Eq:MF-X-PF:analytic:chi:p:q:s}
\end{equation}

Comparing Eq.(\ref{Eq:MF-X-PF:pq:tauxy}) and Eq.~(\ref{Eq:MF-X-PF:analytic:chi:p:q:s}), we obtain the joint mass exponent function:
\begin{equation}
% \begin{aligned}
  \tau_{xy}(p,q) =\frac{p\gamma}{2\ln 2}+\tau_y(Q) =\frac{p\gamma}{2\ln 2}-\frac{\ln[p_y^{Q}+(1-p_y)^{Q}]}{\ln{2}}.%\\=\frac{p\gamma}{2\ln 2}-\frac{\ln[p_y^{\beta p/2+q/2}+(1-p_y)^{\beta p/2+q/2}]}{\ln{2}}
% \end{aligned}
  \label{Eq:MF-X-PF:analytic:tauxy:p:q}
\end{equation}
It follows that
\begin{equation}
% \begin{aligned}
  \alpha_x =\frac{2\partial \tau_{xy}(p,q)}{\partial p}%\\
%  &=\frac{\gamma}{\ln 2}-\frac{\beta}{\ln 2}\frac{p_y^{\beta p/2+q/2}\ln p_y+(1-p_y)^{\beta p/2+q/2}\ln (1-p_y) }{p_y^{\beta p/2+q/2}+(1-p_y)^{\beta p/2+q/2}}\\
  =\frac{\gamma}{\ln 2}-\frac{\beta}{\ln 2}\frac{p_y^{Q}\ln p_y+(1-p_y)^{Q}\ln (1-p_y) }{p_y^{Q}+(1-p_y)^{Q}}
% \end{aligned}
  \label{Eq:MF-X-PF:analytic:alphax}
\end{equation}
and
\begin{equation}
 \begin{aligned}
  \alpha_y = \frac{2\partial \tau_{xy}(p,q)}{\partial q}%\\
%  &=-\frac{1}{\ln 2}\frac{p_y^{\beta p/2+q/2}\ln p_y+(1-p_y)^{\beta p/2+q/2}\ln (1-p_y) }{p_y^{\beta p/2+q/2}+(1-p_y)^{\beta p/2+q/2}}\\
  =-\frac{1}{\ln 2}\frac{p_y^{Q}\ln p_y+(1-p_y)^{Q}\ln (1-p_y) }{p_y^{Q}+(1-p_y)^{Q}}.
 \end{aligned}
  \label{Eq:MF-X-PF:analytic:alphay}
\end{equation}
We obtain immediately the relationship between $\alpha_x$ and $\alpha_y$
\begin{equation}
  \alpha_x =\frac{\gamma}{\ln 2}+\beta\alpha_y.
  \label{Eq:MF-X-PF:analytic:alphaxalphay}
\end{equation}
This relationship explains the observation in Fig.~\ref{Fig:MF-X-PF:pq:pmodel}(i) that $f_{xy}(\alpha_x,\alpha_y)$ is a curve along this line rather than a surface and the line segment (\ref{Eq:MF-X-PF:analytic:alphaxalphay}) is the projection of $f_{xy}(\alpha_x,\alpha_y)$ onto the $(\alpha_x,\alpha_y)$ plane.

We now derive the main geometric properties of $\alpha_x(Q)$ and $\alpha_y(Q)$. We find that $\alpha_y(Q)$ is a monotonically decreasing function of $Q$, because
\begin{equation}
 \frac{d \alpha_y}{d Q} = -\frac{1}{\ln 2}\frac{p_y^{Q}(1-p_y)^{Q}\left[\ln p_y-\ln (1-p_y)\right]^2} {\left[p_y^{Q}+(1-p_y)^{Q}\right]^2} < 0.
  \label{Eq:MF-X-PF:properties:analytic:d_alphay}
\end{equation}
We can prove that the limits of $\alpha_y$ exist when $Q\to\pm\infty$. We rewrite Eq.~(\ref{Eq:MF-X-PF:analytic:alphay}) as follows
\begin{equation}
  \alpha_y  =-\frac{1}{\ln 2}\frac{\ln p_y+\left[(1-p_y)/{p_y}\right]^{Q}\ln (1-p_y) }{1+\left[(1-p_y)/{p_y}\right]^{Q}}.
  \label{Eq:MF-X-PF:properties:analytic:alphay}
\end{equation}
%%
%When $1-p_y>p_y$ or $p_y<0.5$, we have
%%
%\begin{equation}
%     \left\{
%    \begin{aligned}
%      \alpha_{y,\min} &= \lim_{Q\rightarrow  \infty} \alpha_y = -\frac{\ln(1-p_y)}{\ln 2}\\
%      \alpha_{y,\max} &= \lim_{Q\rightarrow -\infty} \alpha_y = -\frac{\ln p_y}{\ln 2}
%    \end{aligned}
%    \right.
%    \label{Eq:MF-X-PF:properties:analytic:minalpha}
%\end{equation}
%%
%When $1-p_y<p_y$ or $p_y>0.5$, we have
%%
%\begin{equation}
%     \left\{
%    \begin{aligned}
%      \alpha_{y,\min} &= \lim_{Q\rightarrow \infty} \alpha_y = -\frac{\ln p_y}{\ln 2} \\
%      \alpha_{y,\max} &= \lim_{Q\rightarrow -\infty} \alpha_y = -\frac{\ln(1-p_y)}{\ln 2}
%    \end{aligned}
%    \right.
%    \label{Eq:MF-X-PF:properties:analytic:maxalpha}
%\end{equation}
%%
%Writing in a different form, we have
%
We can obtain that
\begin{equation}
     \left\{
    \begin{aligned}
      \alpha_{y,\min} &= \lim_{Q\rightarrow \infty} \alpha_y = \min\left\{-\frac{\ln p_y}{\ln 2},-\frac{\ln(1-p_y)}{\ln 2}\right\} \\
      \alpha_{y,\max} &= \lim_{Q\rightarrow -\infty} \alpha_y = \max\left\{-\frac{\ln p_y}{\ln 2},-\frac{\ln(1-p_y)}{\ln 2}\right\}
    \end{aligned}
    \right.
    \label{Eq:MF-X-PF:properties:analytic:max:min:alphay}
\end{equation}
%
%\begin{equation}
%     \lim_{Q\rightarrow \infty} \alpha_y
%    =\left\{
%    \begin{aligned}
%       &-\frac{\ln 1-p_y}{\ln 2},   & 1-p_y>p_y;\\
%       &-\frac{\ln p_y}{\ln 2},      & 1-p_y<p_y.
%    \end{aligned}
%    \right.
%    \label{Eq:MF-X-PF:properties:analytic:minalphsa}
%\end{equation}
%
%
%\begin{equation}
%     \lim_{Q\rightarrow -\infty} \alpha_y
%    =\left\{
%    \begin{aligned}
%       &-\frac{\ln p_y}{\ln 2},   & 1-p_y>p_y;\\
%       &-\frac{\ln 1-p_y}{\ln 2},      & 1-p_y<p_y.
%    \end{aligned}
%    \right.
%    \label{Eq:MF-X-PF:properties:analytic:maxalphsa}
%\end{equation}
%
Therefore, the solution of Eq.~(\ref{Eq:MF-X-PF:analytic:alphay}) exists and is unique if and only if $\alpha_y\in[\alpha_{y,\min},\alpha_{y,\min}]$. The explicit form of the solution is
\begin{equation}
 \begin{aligned}
  Q = {\ln \left[-\frac{\log_2[(1-p_y)/p_y]}{\alpha_y+\log_2 p_y}-1\right]}/{\ln\left[\frac{p_y}{(1-p_y)}\right]}.
 \end{aligned}
 \label{Eq:MF-X-PF:analytic:cpqalphay}
\end{equation}
Further, the width of the singularity spectrum of $\alpha_y$ is
\begin{equation}
  \Delta \alpha_y =\frac{\left|\ln (1-p_y) - \ln p_y\right|}{\ln 2}.
  \label{Eq:MF-X-PF:properties:analytic:deltaalphay}
\end{equation}
These results explain the parallel observation of the contour lines in Fig.~\ref{Fig:MF-X-PF:pq:pmodel}(g). When $p_y=0.5$, $\Delta \alpha_y=0$. In this case, the measure is neither multifractal nor monofractal since it is uniformly distributed on the support.

According to Eq.~(\ref{Eq:MF-X-PF:analytic:alphaxalphay}), we have
\begin{equation}
  \frac{d \alpha_x}{d Q} = -\frac{\beta}{\ln 2}\frac{p_y^{Q}(1-p_y)^{Q}\left[\ln p_y-\ln (1-p_y)\right]^2} {\left[p_y^{Q}+(1-p_y)^{Q}\right]^2},
  \label{Eq:MF-X-PF:properties:analytic:d_alphax}
\end{equation}
which suggests that $\alpha_x$ is a strictly monotonic function of $Q$. Moreover, it is easy to show that
\begin{equation}
     \left\{
    \begin{aligned}
      \alpha_{x,\min} &= \lim_{Q\rightarrow \infty} \alpha_x = \min\left\{-\frac{\ln p_x}{\ln 2},-\frac{\ln(1-p_x)}{\ln 2}\right\} \\
      \alpha_{x,\max} &= \lim_{Q\rightarrow -\infty} \alpha_x = \max\left\{-\frac{\ln p_x}{\ln 2},-\frac{\ln(1-p_x)}{\ln 2}\right\}
    \end{aligned}
    \right.
    \label{Eq:MF-X-PF:properties:analytic:max:min:alphax}
\end{equation}
Therefore, the solution of Eq.~(\ref{Eq:MF-X-PF:analytic:alphax}) exists, which is unique if and only if $\alpha_x\in[\alpha_{x,\min},\alpha_{x,\max}]$. Due to the symmetry between the two measures $m_x$ and $m_y$, the results for $\alpha_x$ are obvious, provided that we know the geometric properties of $\alpha_y$.

We now turn to investigate the geometric properties of the multifractal spectrum $f_{xy}(\alpha_x,\alpha_y)$, which has the following form:
\begin{equation}
 \begin{aligned}
  f_{xy}(\alpha_x,\alpha_y) &= p\alpha_x/2+q\alpha_y/2 -\tau_{xy}(p,q)\\
    & = \frac{p}{2}\left(\frac{\gamma}{\ln 2}+\beta\alpha_y\right)+\frac{q}{2}\alpha_y -\frac{p\gamma}{2\ln 2}+\frac{\ln[p_y^{Q}+(1-p_y)^{Q}]}{\ln{2}}\\
    &=-\frac{Q}{\ln2}\frac{\ln p_y+\left(\frac{1-p_y}{p_y}\right)^{Q}\ln (1-p_y) }{1+\left(\frac{1-p_y}{p_y}\right)^{Q}}
  +\frac{\ln p_y^{Q}+\ln\left[1+\left(\frac{1-p_y}{p_y}\right)^{Q}\right]}{\ln{2}}\\
 %     &=-\frac{1}{\ln 2}\frac{Q\ln p_y+Q\left[\frac{1-p_y}{p_y}\right]^{Q}\ln (1-p_y) }{1+\left[\frac{1-p_y}{p_y}\right]^{Q}}
%  +\frac{1}{\ln 2}\left[Q\ln p_y+\ln [1+\left[\frac{1-p_y}{p_y}\right]^{Q}]\right]\\
      &=\frac{1}{\ln 2}\frac{Q\left(\frac{1-p_y}{p_y}\right)^{Q} \ln\left(\frac{p_y}{1-p_y}\right) +\left[1+\left(\frac{1-p_y}{p_y}\right)^{Q}\right]\ln\left[1+\left(\frac{1-p_y}{p_y}\right)^{Q}\right]} {1+\left(\frac{1-p_y}{p_y}\right)^{Q}}
 \end{aligned}
  \label{Eq:MF-X-PF:analytic:fpq}
\end{equation}
It is easy to find that
\begin{equation}
  f_{xy}(Q=0) = 1 ~~~~~{\rm{and}}~~~~~f_{xy}(Q)=f_{xy}(-Q),
\end{equation}
where $f_{xy}(Q)\triangleq f_{xy}(\alpha_x,\alpha_y; Q)$. It indicates that $f_{xy}(\alpha_x,\alpha_y)$ is symmetric with respect to the line $Q=0$, as numerically shown in Fig.~\ref{Fig:MF-X-PF:pq:pmodel}(h). Furthermore, we obtain
\begin{equation}
  \lim_{Q\to\pm\infty} f_{xy}(p,q) = 0.
  \label{Eq:MF-X-PF:properties:analytic:fpq:infty}
\end{equation}

Taking derivative of $f_{xy}(\alpha_x,\alpha_y)$ with respect to $Q$, we have
\begin{equation}
 \frac{d f_{xy}(Q)}{d Q} = -\frac{Q}{\ln2} \left[\ln\left(\frac{p_y}{(1-p_y)}\right)\right]^2
 {\left[\left(\frac{p_y}{1-p_y}\right)^{{Q}/2}+\left(\frac{1-p_y}{p_y}\right)^{Q/2}\right]^{-2}}.
  \label{Eq:MF-X-PF:properties:analytic:df}
\end{equation}
When $Q<0$, $df_{xy}(Q)/dQ>0$ so that $f_{xy}(Q)$ is a monotonically increasing function of $Q$. When $Q>0$, $df_{xy}(Q)/dQ<0$ so that $f_{xy}(Q)$ is a monotonically decreasing function of $Q$. Therefore, the maximum of $f_{xy}(Q)$ is 1 and its minimum is 0. These properties explain the parallel feature of the contour lines in Fig.~\ref{Fig:MF-X-PF:pq:pmodel}(h).

We note that the numerical results are in excellent agreement with the analytical results for $\tau_{xy}(p,q)$ in Eq.~(\ref{Eq:MF-X-PF:analytic:tauxy:p:q}), $\alpha_x(p,q)$ in Eq.~(\ref{Eq:MF-X-PF:analytic:alphax}), $\alpha_y(p,q)$ in Eq.~(\ref{Eq:MF-X-PF:analytic:alphay}), and $f_{xy}(p,q)$ in Eq.~(\ref{Eq:MF-X-PF:analytic:fpq}). Combining Eq.~(\ref{Eq:MF-X-PF:analytic:cpqalphay}) and Eq.~(\ref{Eq:MF-X-PF:analytic:fpq}), we find that $f_{xy}(\alpha_x,\alpha_y)$ is a univariate function of $\alpha_y$, or of $\alpha_x$ by using Eq.~(\ref{Eq:MF-X-PF:analytic:alphaxalphay}).

\subsection{Numerical analysis applying MF-X-PF$(q)$}

\begin{figure}[htb]
  \centering%
\includegraphics[width=0.96\linewidth]{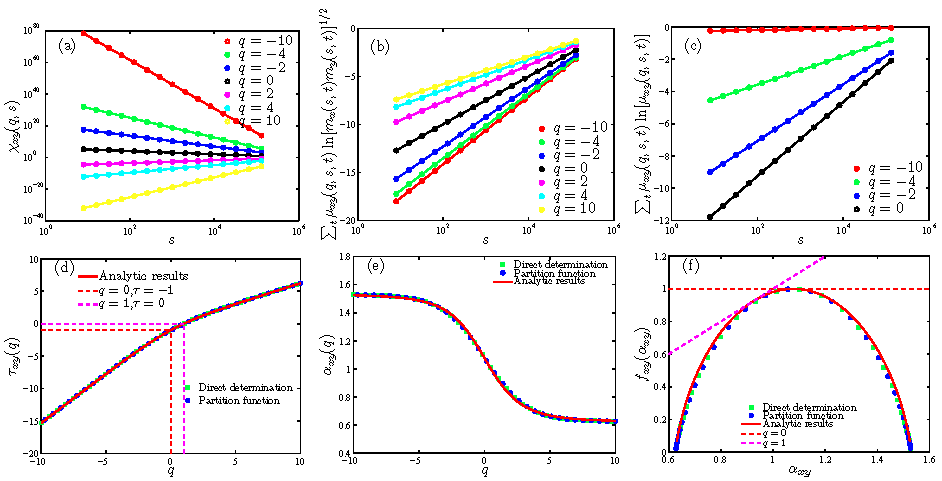}
  \caption{Joint multifractal analysis of two binomial measures with $p_x=0.3$ and $p_y=0.4$ based on the MF-X-PF$(q)$ method. (a) Power-law dependence of $\chi_{xy}(q,s)$ on box size $s$ for different $q$. (b) Linear dependence of $\sum_t \mu_{xy}(q,s,t) \ln{[m_{x}(s,t)m_y(s,t)]^{1/2}}$ against $\ln{s}$. (c) Linear dependence of $\sum_t\mu_{xy}(q,s,t)\ln[\mu_{xy}(q,s,t)]$ against $\ln{s}$. (d) The mass exponent function $\tau_{xy}(q)$. (e) The singularity strength function $\alpha(q)$. (f) The multifractal singularity spectrum $f_{xy}(\alpha)$. }
  \label{Fig:MF-X-PF:p=q:p-model}
\end{figure}

We also apply the MF-X-PF$(q)$ method to the same mathematical example. The results are shown in Fig.~\ref{Fig:MF-X-PF:p=q:p-model}. We find that the three theoretical relationships in Eq.~(\ref{Eq:MF-X-PF:q:tauxy:taux:tauy}), Eq.~(\ref{Eq:MF-X-PF:q:alpha:alphax:alphay}), and Eq.~(\ref{Eq:MF-X-PF:q:fxy:fx:fy}) are nicely verified. In addition, we observe again that the results from the classic partition function approach and the direct determination approach agree with each other. We note that this is also the case for other mathematical and empirical examples investigated in this work. Thus we will not show the results obtained from the direct determination approach in the rest of this paper.

\section{Joint multifractal analysis of bivariate fractional Brownian motions}

We further investigate the MF-X-PF$(p,q)$ algorithm using monofractal measures. If $m_x$ and $m_y$ are monofractal, we have $\alpha_x=\alpha_y=1$ and $f_{xy}=1$ according to its definition in Eq.~(\ref{Eq:MF-X-PF:Ns:s:f:alpha:xy}) \cite{Halsey-Jensen-Kadanoff-Procaccia-Shraiman-1986-PRA,Jiang-Zhou-2008a-PA}. Together with Eq.~(\ref{Eq:MF-X-PF:f:alpha}), we have
\begin{equation}
  \tau_{xy}(p,q)=p/2+q/2-1.
  \label{Eq:MF-X-PF:monofractal:tau}
\end{equation}
These properties are indicators of monofractality.

\begin{figure}[htb]
%\centering%
%  \includegraphics[width=0.32\linewidth]{Fig_MF-X-PF_pq_BFBM_chi_s_p2.eps}
%  \includegraphics[width=0.32\linewidth]{Fig_MF-X-PF_pq_BFBM_tau.eps}
%  \includegraphics[width=0.32\linewidth]{Fig_MF-X-PF_pq_BFBM_tau_Error.eps}\\
%  \includegraphics[width=0.32\linewidth]{Fig_MF-X-PF_pq_BFBM_alphax.eps}
%  \includegraphics[width=0.32\linewidth]{Fig_MF-X-PF_pq_BFBM_alphay.eps}
%  \includegraphics[width=0.32\linewidth]{Fig_MF-X-PF_pq_BFBM_f_alphaxy.eps}
%\includegraphics[width=0.96\linewidth]{Fig3.eps}
\includegraphics[width=0.96\linewidth]{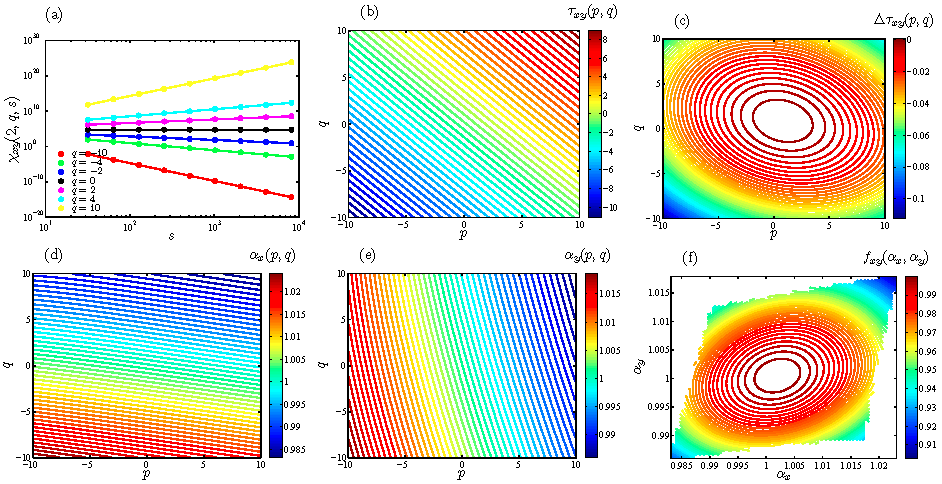}
  \caption{Joint multifractal analysis of bivariate fractional Brownian motions with $H_{xx}=0.1$, $H_{yy}=0.5$ and $\rho=0.5$. (a) Power-law dependence of $\chi_{xy}(p,q,s)$ on box size $s$ for different $q$ with fixed $p=2$. (b) Mass exponent function $\tau_{xy}(p,q)$ function obtained from (a). (c) Errors $\Delta \tau_{xy}(p,q)$ between the estimated exponent $\tau_{xy}(p,q)$ and the theoretical function $p/2+q/2-1$. (d,e,f) Singularity functions $\alpha_x(p,q)$ and $\alpha_y(p,q)$, and multifractal spectra $f_{xy}(\alpha_x, \alpha_y)$ obtained from (b). }
  \label{Fig:MF-X-PF:BFBM}
\end{figure}

The mathematical model used here is bivariate fractional Brownian motions (BFBMs). The two components $x(t)$ and $y(t)$ of the BFBM are two univariate fractional Brownian motions with Hurst indices $H_{xx}$ and $H_{yy}$, respectively. The basic properties of multivariate fractional Brownian motions have been comprehensively studied \cite{Lavancier-Philippe-Surgailis-2009-SPL,Coeurjolly-Amblard-Achard-2010-EUSIPCO,Amblard-Coeurjolly-Lavancier-Philippe-2013-BSMF}. Extensive numerical experiments of other MF-DCCA algorithms have been conducted using bivariate fractional Brownian motions \cite{Jiang-Zhou-2011-PRE,Qian-Liu-Jiang-Podobnik-Zhou-Stanley-2015-PRE}.
The two Hurst indexes $H_{xx}$ and $H_{yy}$ of the two univariate FBMs and their cross-correlation coefficient $\rho$ are input arguments of the simulation algorithm. By using the simulation procedure described in Refs.~\cite{Coeurjolly-Amblard-Achard-2010-EUSIPCO,Amblard-Coeurjolly-Lavancier-Philippe-2013-BSMF}, we have generated as an example a realization of BFBM with $H_{xx}=0.1$, $H_{yy}=0.5$ and $\rho=0.5$. The length of the BFBM is $2^{16}$. The joint multifractal analysis of the BFBM using the MF-X-PF$(p,q)$ algorithm is presented in Fig.~\ref{Fig:MF-X-PF:BFBM}.

The corresponding power-law dependence of the joint partition function $\chi_{xy}(p,q,s)$ with respect to the box size $s$ for different $q$'s and fixed $p=2$ is shown in Fig.~\ref{Fig:MF-X-PF:BFBM}(a). The scaling ranges span over two orders of magnitude. The slopes of the lines give the estimates of $\tau_{xy}(p,q)$, where $p$ and $q$ vary from $-10$ to $10$ with a spacing of 0.1. The resulting mass exponents $\tau_{xy}(p,q)$ are shown in the contour plot of Fig.~\ref{Fig:MF-X-PF:BFBM}(b). We observe that $\tau_{xy}(p,q)$ increases with $p$ and $q$, the contour curves are parallel lines, and the parallel lines are evenly spaced. These features suggest that $\tau_{xy}(p,q)$ is a linear function of $p$ and $q$, which is an indicator of monofractality.

In order to further show the performance of the MFXPF algorithm, we calculate the errors between the estimated exponents $\tau_{xy}(p,q)$ and the theoretical exponents as $\Delta \tau_{xy}(p,q)=\tau_{xy}(p,q)-[p/2+q/2-1]$. Fig.~\ref{Fig:MF-X-PF:BFBM}(c) shows the dependence of $\Delta \tau_{xy}(p,q)$ with respect to $p$ and $q$. All the $\Delta \tau_{xy}(p,q)$ values are less than 0.15, implying that the algorithm gives good estimates.

By adopting the double Legendre transform in Eq.~(\ref{Eq:MF-X-PF:pq:Legendre}) numerically, we get the singularity strength functions $\alpha_x(p,q)$ and $\alpha_y(p,q)$ and the multifractal spectrum $f_{xy}(\alpha_x, \alpha_y)$, whose contour plots are shown in Fig.~\ref{Fig:MF-X-PF:BFBM}(d,e,f). The singularity strength functions $\alpha_x(p,q)$ and $\alpha_y(p,q)$ are close to 1, indicating that the functions $\alpha_x(p,q)$ and $\alpha_y(p,q)$ are independent of the order $p$ and $q$. Although there is a trend in each function $\alpha_x(p,q)$ and $\alpha_y(p,q)$, the theoretical functions $\alpha_x(p,q)=1$ and $\alpha_y(p,q)=1$ are basically satisfied. Hence, the MF-X-PF algorithm is able to correctly capture the monofractal nature of the BFBMs.

Fig.~\ref{Fig:MF-X-PF:BFBM}(g) plots the singularity spectrum $f_{xy}(\alpha_x, \alpha_y)$, which is a surface and the contour lines are closed curves. It is easy to find that the vast majority of the surface is nearly equal to the theoretical function $f_{xy}(\alpha_x, \alpha_y)=1$. We observe that the errors $\Delta \tau_{xy}(p,q)$ is equal to the difference between $f_{xy}(p,q)$ and 1, as shown by the Legendre transform.

We point out that the results using the direct determination approach are exactly the same as shown in Fig.~\ref{Fig:MF-X-PF:BFBM}. We thus summarize that the theoretical analysis is well verified by the numerical results.

\section{Application to stock market indexes}

We now apply the MF-X-PF$(p,q)$ algorithm to investigate the long-range power-law cross correlations of the daily volatility time series of the Dow Jones Industrial Average (DJIA) and the National Association of Securities Dealers Automated Quotations (NASDAQ) index. The daily volatility is defined as the absolute value of the logarithmic difference of daily closing prices:
\begin{equation}
  R(t)=|\ln P(t)-\ln P(t-1)|,
  \label{Eq:MF-X-PF:return}
\end{equation}
where $P(t)$ is the closing price on day $t$ and has been retrieved for the DJIA and NASDAQ indices. The time period of the samples is from 5 February 1971 to 25 January 2011, containing 10084 data points. The daily return time series of the two indexes are shown in Figure S1 (New J. Phys. online).

\begin{figure}[htb]
\centering
\includegraphics[width=0.96\linewidth]{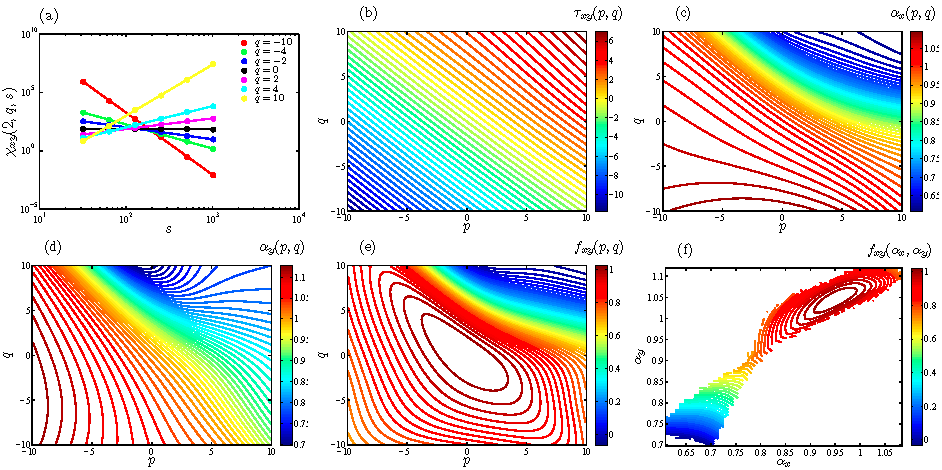}
  \caption{Joint multifractal analysis of the cross correlations between the daily volatility time series of DJIA index and NASDAQ index using the MF-X-PF$(p,q)$ approach. (a) Power-law dependence of $\chi_{xy}(p,q,s)$ on box size $s$ for different $q$'s with fixed $p=2$. (b) Mass exponent function $\tau_{xy}(p,q)$. (c) Singularity strength function $\alpha_x(p,q)$. (d) Singularity strength function $\alpha_y(p,q)$. (e) Multifractal function $f_{xy}(p,q)$. (f) Multifractal singularity spectrum $f_{xy}(\alpha_x,\alpha_y)$. }
  \label{Fig:MF-X-PF:Index}
\end{figure}

Fig.~\ref{Fig:MF-X-PF:Index}(a) shows on log-log scales the dependence of the joint partition function $\chi_{xy}(p,q,s)$ with respect to the box size $s$ for different $q$'s and fixed $p=2$. We observe nice power-law scaling over about 1.5 orders of magnitude. The contour plot of the exponents $\tau_{xy}(p,q)$ is shown in Fig.~\ref{Fig:MF-X-PF:Index}(b), where $p$ and $q$ vary from $-10$ to $10$ with a spacing of 0.1. The contour curves are not straight lines and the spacings between neighboring curves are not equidistant. Fig.~\ref{Fig:MF-X-PF:Index}(c) and Fig.~\ref{Fig:MF-X-PF:Index}(d) illustrate respectively the contour plots of the singularity strength functions $\alpha_x(p,q)$ and $\alpha_y(p,q)$, which are obtained numerically from $\tau_{xy}(p,q)$. We observe that the values of the singularity strength range from 0.6 to 1.2, which are well dispersed. In addition, the singularity strength functions are not monotonic with respect to $p$ or $q$. Fig.~\ref{Fig:MF-X-PF:Index}(e) illustrates the multifractal function $f_{xy}(p,q)$ obtained from the Legendre transform, whose values range from 0 to 1. The maximum $f_{xy}(p,q)=1$ is reached at point $(p,q)=(0,0)$. Within the investigated intervals of $p$ and $q$, the small $f_{xy}(p,q)$ values concentrated in the region with large values of $p$ and $q$. In Fig.~\ref{Fig:MF-X-PF:Index}(f), we present the singularity spectrum $f_{xy}(\alpha_x,\alpha_y)$. These empirical findings suggest that the cross correlations between daily volatilities of DJIA and NASDAQ possess multifractal nature, which is consistent with previous results using the MF-X-DFA, MF-X-DMA and MF-X-PF$(q)$ methods \cite{Zhou-2008-PRE,Jiang-Zhou-2011-PRE,Wang-Shang-Ge-2012-Fractals,Xiong-Shang-2015-CNSNS}.

\begin{figure}[htb]
  \centering
\includegraphics[width=0.96\linewidth]{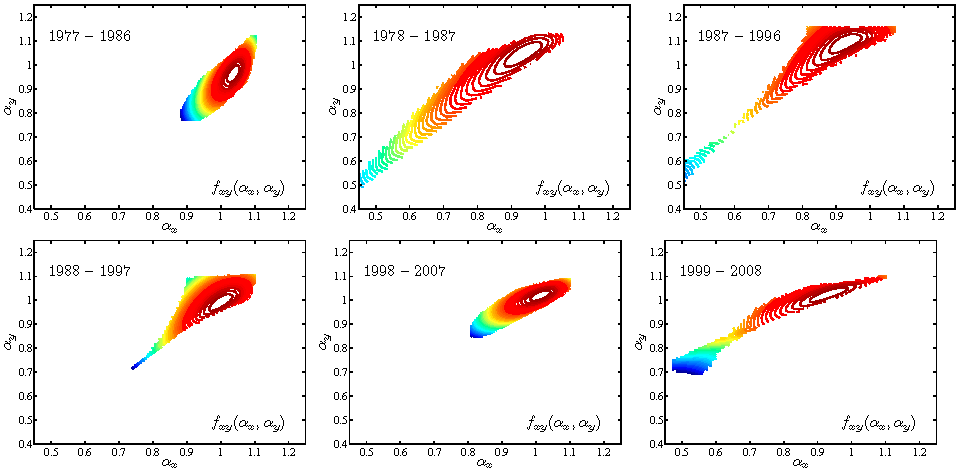}
  \caption{Comparison of the joint multifractal singularity spectrum $f_{xy}(\alpha_x,\alpha_y)$ between the daily volatility time series of the DJIA index and the NASDAQ index in different time periods with and without market turmoil.}
  \label{Fig:MF:X:PF:index:t}
\end{figure}

To reveal whether the joint multifractality between the daily volatilities of the two indices remains or changes along time, we perform the MF-X-PF$(p,q)$ analysis in moving windows on a decade basis with a step of one year. The results are presented in Figure S2 (New J. Phys. online). We show six plots in Fig.~\ref{Fig:MF:X:PF:index:t}. We find that the joint multifractal singularity spectrum $f_{xy}(\alpha_x,\alpha_y)$ changes over time. Moreover, the inclusion or exclusion of financial turmoils (high volatile periods) has a significant impact on the shape of $f_{xy}(\alpha_x,\alpha_y)$. In the sample period under investigation, there were two infamous market crises, the Black Monday in 1987 and the latest crisis in 2008. During relatively calm periods, the $f_{xy}(\alpha_x,\alpha_y)$ contour looks roughly like an American football. However, when one of the crisis is included, the contours are significantly stretched to the southwest. In other words, the singularity strengthes $\alpha_x$ and $\alpha_y$ have much smaller values during turmoil periods. This is actually not surprising because this feature is well-documented for ordinary multifractals \cite{Halsey-Jensen-Kadanoff-Procaccia-Shraiman-1986-PRA}.

We repeat the same analysis for two stocks Du Pont (NYSE:DD) and Exxon Mobil (NYSE:XOM) over time period from 05-Jan-1970 to 01-Sep-2015, containing 11522 data points. The daily return time series of the two stocks are shown in Figure S3 (New J. Phys. online). The results are illustrated in Figure S4 (New J. Phys. online). As expected, very similar results are observed. Two more pairs of financial time series are investigated and the results are presented in Figure S5 to Figure S8 of the Supplementary data (New J. Phys. online). One pair is about crude oil commodities, Arab Light to USA and WTI Cushing. The sample period from is 03-Jan-1991 to 18-Dec-2012, containing 5510 data points. Another pair is about Special Drawing Rights (SDRs) per
currency unit for the U.K. pound sterling (GBP) and the U.S. dollar (USD) over time period from 05-Jan-1994 to 01-Sep-2015, containing 5452 data points.

Compared with the results of binomial measures and fractional Brownian motions, the multifractal function and the multifractal singularity spectrum exhibit different shapes for different data sets studied. For example, in Fig.~\ref{Fig:MF-X-PF:Index}(f) for the financial market data there is a pronounced asymmetry, and the spectrum exhibits a stretched shape, in sharp contrast to Fig.~\ref{Fig:MF-X-PF:BFBM}(f) for the artificial BFBM data. These features reflect the irregular nonlinear traits of financial indexes. Roughly, the spectrum contour parallels to the diagonal $\alpha_x=\alpha_y$ (cf. Eq.~(\ref{Eq:MF-X-PF:analytic:alphaxalphay})), which is due to the fact that the DJIA and NASDAQ indexes comove along time so that the volatilities fulfill Eq.~(\ref{Eq:MF-X-PF:analytic:mxmy}) to certain extend. A direct conjecture is that the correlation coefficient $\rho(\alpha_x,\alpha_y)$ is greater if the correlation coefficient $\rho(R_x(t),R_y(t))$ is greater. This is validated by Fig.~S9 in the Supplementary Data (New J. Phys. online).

\section{Conclusions}

We have studied the properties of joint multifractal analysis based on partition function with two moment orders, termed MF-X-PF$(p,q)$. The uni-order method MF-X-PF$(q)$ has then been derived. The main properties of these methods have been obtained analytically. For instance, for the MF-X-PF$(q)$ method, we have obtained the relationship between the joint mass exponent function and the individual mass exponent functions, $\tau_{xy}(q)=[\tau_x(q)+\tau_y(q)]/2$, which was numerically and empirically observed in the literature.

We applied the MF-X-PF$(p,q)$ method to multifractal binomial measures. The expressions of mass function, singularity strength and multifractal spectrum of the cross correlations have been derived, which agree excellently with the numerical results. We further validated the performance of the method by using bivariate fractional Brownian motions without multifractal cross correlations. When applied to the daily volatility time series of two stock market indexes, intriguing multifractality in the cross correlations is confirmed. The multifractal properties of these examples are found to be the same when we use the conventional determination approach and the direct determination approach.

Multifractal cross-correlation analysis has been applied in many fields, especially in Econophysics. Although there are numerous methods, most of them consider only one moment order. It is natural that bi-order methods such as MF-X-PF$(p,q)$ can be developed for other uni-order methods. We expect that such bi-order methods will unveil new stylized facts in the analysis of financial time series, which can serve to calibrate agent-based models \cite{Li-Zhang-Zhang-Zhang-Xiong-2014-IS}. In addition, the joint multifractal nature extracted from two long-range cross-correlated time series has potential applications. One possibility is to construct a multi-scale cross-correlation measure, analogous to other DCCA coefficients \cite{Zebende-2011-PA,Podobnik-Jiang-Zhou-Stanley-2011-PRE,Zebende-daSilva-Filho-2013-PA,Kristoufek-2014a-PA,Kristoufek-2014b-PA}. Another possibility is to construct a measure quantifying market efficiency \cite{DiMatteo-Aste-Dacorogna-2005-JBF,Zunino-Tabak-Figliola-Perez-Garavaglia-Rosso-2008-PA,Zunino-Figliola-Tabak-Perez-Garavaglia-Rosso-2009-CSF,Wang-Liu-Gu-2009-IRFA}. A related possibility is to quantitatively characterize the degree of market unrest other than the volatility measure \cite{Oh-Eom-Havlin-Jung-Wang-Stanley-Kim-2012-EPJB}.

\section*{Acknowledgments}

We are grateful to the referees for their insightful suggestions. We acknowledge financial support from the National Natural Science Foundation of China (11375064 and 71131007), the Program for Changjiang Scholars and Innovative Research Team in University (IRT1028), and the Fundamental Research Funds for the Central Universities.

%\section*{References}

%\bibliography{E:/Papers/Auxiliary/Bibliography}

\providecommand{\newblock}{}

\end{document}